\documentclass[notitlepage,twocolumn,prx,nobalancelastpage,amssymb,showpacs]{revtex4-1}

\usepackage[utf8]{inputenc}
\usepackage{amssymb}
\usepackage{amsmath}
\usepackage{color}
\usepackage{graphicx}

\newcommand{\be}{\begin{equation}}
\newcommand{\ee}{\end{equation}}
\newcommand{\bea}{\begin{eqnarray}}
\newcommand{\eea}{\end{eqnarray}}

\newcommand{\Ket}[1]{|#1\rangle}
\newcommand{\Bra}[1]{\langle #1|}

\newcommand{\avg}[1]{{\left\langle #1\right\rangle}}

\renewcommand{\vec}[1]{{\bf #1}}
\renewcommand{\epsilon}{\varepsilon}

\begin{document}
\title{
Achieving quantized transport in Floquet topological insulators via energy filters}
\author{Ruoyu Zhang$^1$, Frederik Nathan$^2$, Netanel H. Lindner$^3$, and Mark S. Rudner$^1$}
\affiliation{
        $^1$Department of Physics, University of Washington, Seattle, Washington 98195, USA
        \\
        $^2$Niels Bohr Institute, Copenhagen University, 2100 Copenhagen, Denmark
        \\
        $^3$Physics Department, Technion, 3200003 Haifa, Israel
}

\begin{abstract}
    
    Due to photon-assisted transport processes, 
    chiral edge modes induced by periodic driving do not directly mediate quantized transport. 
    Here we show how narrow bandwidth ``energy filters'' can restore quantization by suppressing photon assisted transport through Floquet sidebands.
    We derive a Floquet Landauer type equation to describe transport through such an energy-filtered setup, and show how the filter can be integrated out to yield a sharply energy-dependent renormalized system-lead coupling.
    We show analytically and through numerical simulations that a nearly quantized conductance can be achieved in both off-resonantly and resonantly induced quasienergy gaps when filters are introduced. 
    The conductance approaches the appropriate quantized value on each plateau with increasing system and filter size. 
    We introduce a ``Floquet distribution function'' and show both analytically and numerically that it approaches the equilibrium Fermi-Dirac form when narrow-band filters are introduced, highlighting the mechanism that restores quantized transport.
\end{abstract}

\maketitle

The goal of Floquet engineering is to endow a physical system with new properties or functionalities ``on demand'' through the application of time-periodic driving \cite{Basov2017}. 
The tantalizing possibility of dynamically inducing robust topological phenomena in otherwise trivial systems has been a subject of particular interest in recent years \cite{Eckardt2017, Oka2019, RudnerLindnerReview, Cooper2019, Ozawa2019, Harper2019}.
Through a number of recent experiments, topologically nontrivial Floquet-Bloch bands have been created and imaged in cold atomic systems \cite{Jotzu2014, Aidelsburger2015, Flaschner2016, Maczewsky2017, Mukherjee2017, Wintersperger2020, Braun2023}, and signatures such as dynamically-induced band gaps and an optically-induced Hall effect have been observed in solid state systems \cite{Wang2013, McIver2018, Zhou23}.

\begin{figure}[h!]
	\includegraphics[width=\columnwidth]{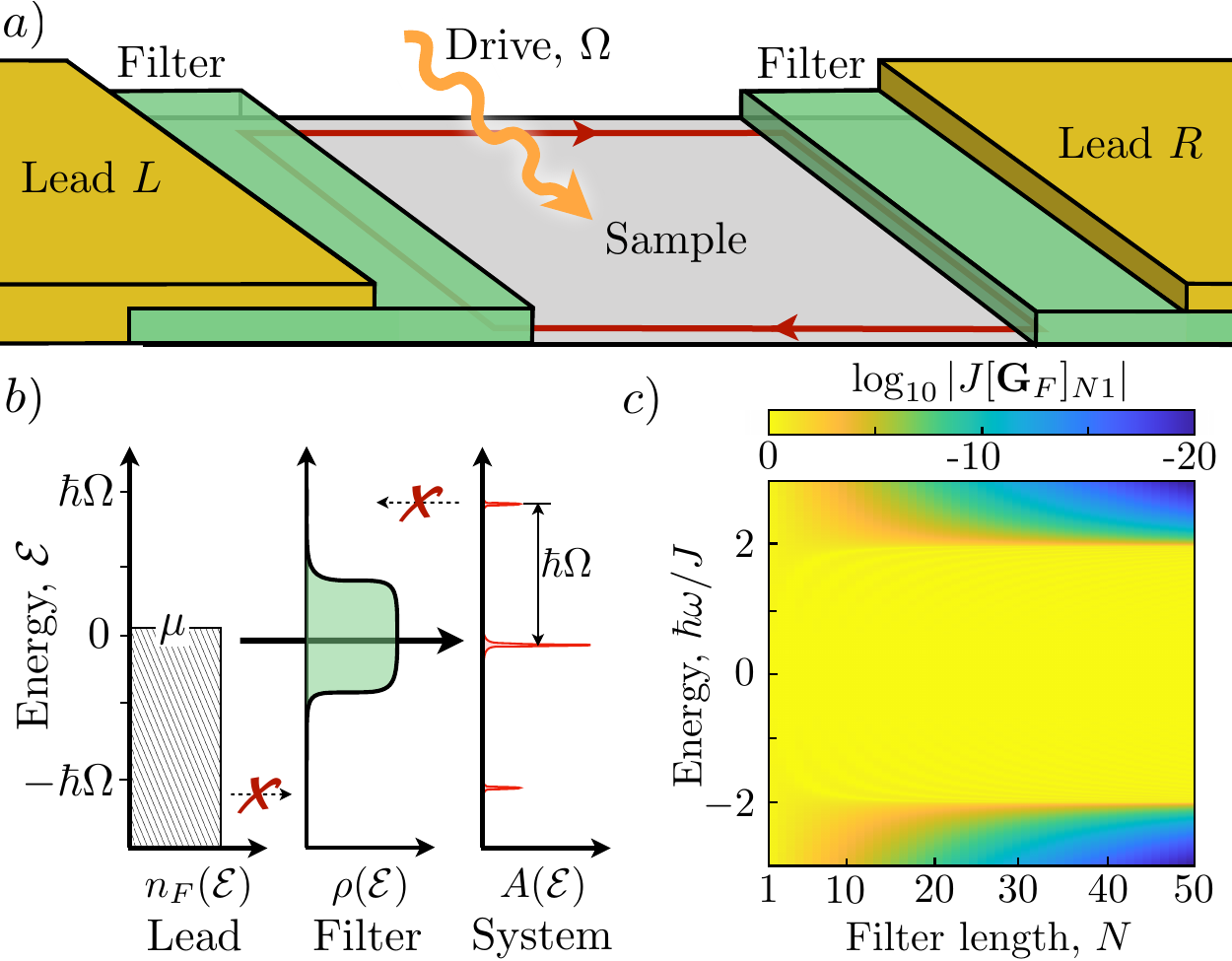} 
	\caption{Energy-filtered leads for achieving quantized transport through Floquet edge modes.
	$a$) Two ``energy filters'' are placed between the metallic leads and the system.
	Each energy filter is a short segment of a material hosting a narrow band (of bandwidth less than $\hbar\Omega$, where $\Omega$ is the drive angular frequency).
	 The energy filters suppress photon-assisted transport processes that generally violate the conditions under which quantized transport is expected.
	 $b)$ The narrow filter density of states $\rho(\mathcal{E})$ allows coupling between the lead and a single peak of the Floquet edge state's spectral distribution. The coupling to sidebands outside the filter energy window is strongly suppressed.
	$c$) Energy selectivity of the filter as a function of filter length, for an $N$ site one-dimensional tight-binding chain with hopping parameter $J$.
We plot the filter Green's function component  $[{\bf G}_{\rm F}(\omega)]_{L_{\rm F}1}$ that describes propagation of a particle from one end of the filter to the other (as it transits from the lead to the system), see Eqs.~(\ref{eq:FloquetLandauer}) and (\ref{eq:filterGamma}).
Large values of $|[{\bf G}_{\rm F}(\omega)]_{L_{\rm F}1}|$ correspond to high transmission probabilities through the filter.}
\label{fig:filter_sketch}
\end{figure}
A key signature of topologically nontrivial Bloch bands is the appearance of robust edge or surface modes at system boundaries or interfaces where topological indices differ.
In equilibrium, such modes are 
expected to facilitate various types of quantized transport \cite{ThoulessPump, TKNN, Halperin1982, Buttiker88}. Under the assumption of good contacts between the leads and the sample edges, the Fermi distribution of the source is mapped directly onto the chiral edge mode, which gives rise to a quantized conductance $G_{\rm H} \equiv e^2/h$.
The situation is more subtle for systems with dynamically-induced topological ``Floquet'' edge modes~\cite{Kundu2014,FoaTorresMultiTerminal, Farrell2015, Farrell2016, Mueller2020}, which mediate photon-assisted transport processes that break the quantization of conductance. 

Topological edge modes of periodically driven systems are composed of bands of single-particle {\it Floquet states} \cite{Eckardt2017, Cooper2019, Oka2019, RudnerLindnerReview}.
Importantly, the spectral weight of each Floquet state is spread across many sidebands, which are separated in energy by integer multiples of the driving field photon energy $\hbar \Omega$, where $\Omega$ is the angular frequency of the drive. 
Through these sidebands, each Floquet state within the edge mode couples to states at many energies within the non-driven lead, both above and below the lead's Fermi energy.
%
Consequently, the Fermi distributions of the leads are not simply mapped onto the edge states, and the differential conductance generically differs from the quantized value $e^2/h$.

\begin{figure}[t]
	\includegraphics[width=\columnwidth]{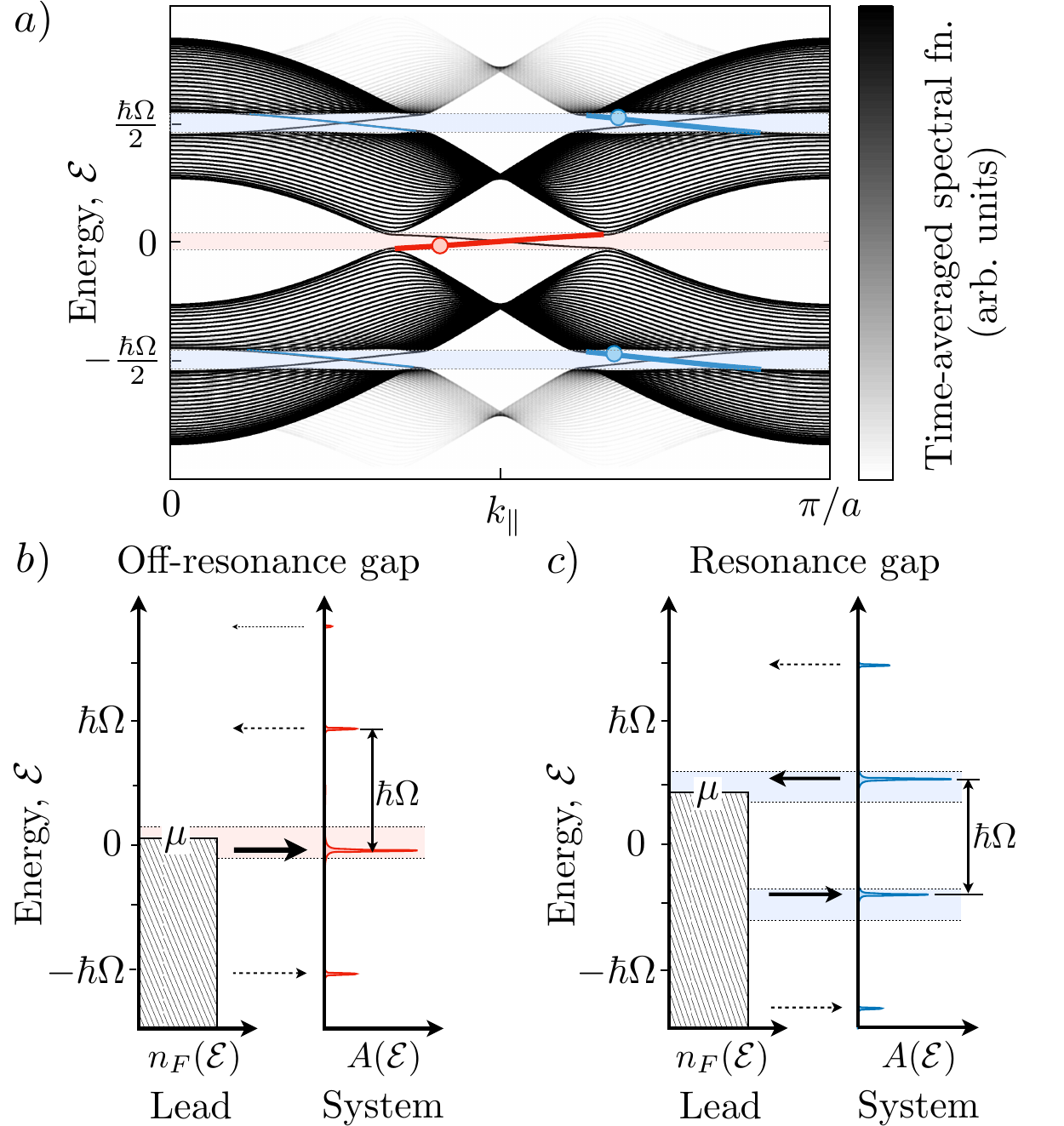} 
	\caption{Transport through Floquet edge modes.
	$a$) Time-averaged spectral function of a graphene-like system subjected to circularly polarized light, showing drive-induced chiral edge modes both in the gap that opens near the Dirac point and in the resonance gap where the drive resonantly couples states in the valence and conduction bands. The spectral weight of each state of the Floquet edge mode is spread across several sidebands, spaced by integer multiples of the driving field photon energy $\hbar\Omega$.
	$b, c$) When the system is connected to leads, each state of the edge mode couples to several states of the leads via the sidebands exhibited in the time-averaged spectral function $A(\mathcal{E})$.
	This prevents a sharp Fermi surface from being established in the Floquet edge mode, and leads to deviations from the quantized conductance that is expected for chiral edge modes in non-driven systems.
	Panel $b)$ illustrates the situation for chiral edge modes in the off-resonantly induced Floquet gap around the original Dirac point (at energy $\mathcal{E} = 0$), while panel $c)$ illustrates the situation for the chiral edge modes in the resonance-induced Floquet gap, where valence and conduction band states are strongly hybridized.}
\label{fig:summary}
\end{figure}
In this work we show how quantized transport can be restored for drive-induced toplogical systems through the use of energy-filtered leads.
As depicted in Fig.~\ref{fig:filter_sketch}a, the energy filter (green color) is a finite length segment of material placed between the metallic contact and the system.
The filter ideally hosts a well isolated, narrow band of states (of bandwidth less than $\hbar \Omega$) centered in energy near the chemical potential of the contact [see filter density of states $\rho(
\mathcal{E})$ in Fig.~\ref{fig:filter_sketch}b].
Due to the filter's narrow bandwidth, each Floquet state in the system effectively couples to the lead only through a {\it single} sideband, at a well-defined energy. 
Consequently, 
the Fermi distributions of the leads can be mapped directly onto the {\it Floquet edge states}, yielding quantized transport through the edge states. 
In contrast to previous works which incorporated idealized narrow band leads into numerical studies of ballistic transport~\cite{Yap2017} 
and steady state transport in the presence of driving and dissipation~\cite{Esin2018}, 
here we provide an analytical characterization of energy-filtered transport, including explicit modeling of the filters and a systematic study of their restoration of quantization {\it without} the application of additional sum rules (cf.~Refs.~\cite{Kundu2014, Farrell2015, Yap2017}). 

Below we begin by describing the general setup. We first derive a Landauer-type formula for transport through a periodically-driven system coupled to wide-band leads via narrow-band energy filters.
We show how the action of the filters can be characterized via the energy-dependence of the transmission matrices that describe coupling between the system and the leads.
We then present numerical results for a graphene-like (honeycomb tight-binding) system subjected to a circularly polarized driving field~\cite{Oka2009, Kitagawa2011, Gu2011, Usaj2014}.
These numerical results explicitly demonstrate how the introduction of energy filters can promote quantized transport through edge states in both resonance-induced and off-resonance Floquet gaps.

{\it Floquet-Landauer transport through filtered leads.---}
The setup consists of five elements connected in series: the left lead, the left filter, the system, the right filter, and the right lead (Fig.~\ref{fig:filter_sketch}a).
The corresponding Hamiltonian thus consists of several parts:
$ H(t) = H_{\rm S}(t) + \sum_{\lambda}\left[ H_{\rm L,\lambda}+  H_{\rm F,\lambda} + H_{\rm SF,\lambda} + H_{\rm FL,\lambda} \right]$, where  $\lambda =\{\ell, r\}$ labels left and right.
Here $H_{\rm S}(t)$, $H_{\rm L,\lambda}$, and $H_{\rm F,\lambda}$ act on the system, lead $\lambda$, and filter $\lambda$, respectively, while 
$H_{\rm SF,\lambda}$ and $H_{\rm FL,\lambda}$ describe corresponding system-filter and filter-lead couplings. 
Throughout this work we assume that the periodic drive with period $T$ (and corresponding angular frequency $\Omega \equiv 2 \pi / T$) only acts within the system.
The filter, lead, and coupling Hamiltonians are all time-independent. 

In second quantized form, we write the system, lead, and filter Hamiltonians as: 
$H_{\rm S}(t) = {\bf c}^\dagger {\bf H}_{\rm S}(t) {\bf c}$, 
$H_{\rm L,\lambda} = \vec{a}_\lambda^\dagger {\bf H}_{\rm L,\lambda}\vec{a}_\lambda$, and $H_{\rm F,\lambda} = \vec{f}_\lambda^\dagger {\bf H}_{\rm F,\lambda}\vec{f}_\lambda$.
Here $\vec{c} = ( \cdots c_i \cdots)^T$ consists of a complete set of fermionic annihilation operators 
for the system. The vectors $\vec{f}_\lambda$ and $\vec{a}_\lambda$ are defined similarly for the filters and the leads respectively.
With these definitions, the coupling terms become 
$H_{\rm SF,\lambda} = {\bf c}^\dagger {\bf H}_{\rm SF,\lambda} {\bf f}_{\lambda} + {\rm h.c.}$,
and
$H_{\rm FL,\lambda} = {\bf f}_\lambda^\dagger {\bf H}_{\rm FL,\lambda} {\bf a}_{\lambda} + {\rm h.c.}$. 

As detailed in Appendix \ref{app:Landauer}, we derive a Landauer-type formula for the current through the energy-filtered driven system using the equation of motion approach (see, e.g., Ref.~\onlinecite{Kohler2005}). 
We begin by expressing the net current flowing out of the system to the right as the rate of change of the number of particles in the right filter and right lead: $I(t) = \frac{d}{dt}(N_{{\rm F},r} + N_{{\rm L},r})$, where $N_{{\rm F},r} = \vec{f}^\dagger_r \vec{f}_r$ and $N_{{\rm L},r} = \vec{a}^\dagger_r \vec{a}_r$. 
We assume that the setup was initialized at a time $t_0 < 0$ in the distant past, such that it has reached a time-periodic steady state by time $t = 0$. 
The period-averaged steady state current that flows through the system is thus given by 
$\bar{I} =
\frac{1}{T} \int_0^T dt\, \frac{2e}{\hbar} {\rm Im} \avg{{\bf c}^\dagger(t){\bf H}_{{\rm SF},r} {\bf f}_r(t)}$.

To evaluate $\bar{I}$, 
we assume that at the initial time $t_0$ each lead $\lambda$ was in an equilibrium state described by the chemical potential $\mu_\lambda$ and inverse temperature $\beta_\lambda$. 
This condition is encoded mathematically as 
$\langle a_{\lambda\nu}^\dagger(t_0) a_{\lambda'\nu'}(t_0)\rangle=\delta_{\lambda\lambda'}\delta_{\nu\nu'} n_{\lambda}(\epsilon_{\lambda\nu})$, where $n_{\lambda}$ is the Fermi distribution for lead $\lambda$ and $\epsilon_{\lambda\nu}$ is the energy of single particle eigenstate $\nu$ of $H_{\rm L,\lambda}$. 
%
By formally 
integrating the Heisenberg equations of motion for 
$\vec{a}_\lambda(t)$, $\vec{f}_\lambda(t)$ and $\vec{c}(t)$, we express $\bar{I}$ in terms of an expectation value involving only $\vec{a}_{\lambda}(t_0)$ and $\vec{a}_{\lambda}^\dagger(t_0)$, and the single particle Green's functions of the system and the filter (see Appendix \ref{app:Landauer}): $[\vec{G}_{\rm S}(t,t')]_{ij}=-i\theta(t-t')\langle \{{c}_i(t), {c}_j^\dagger(t')\}\rangle$ and $[\vec{G}_{\rm F,\lambda}(t, t')]_{\alpha \beta}=-i\theta(t-t')\langle \{{f}_{\lambda, \alpha}(t), {f}_{\lambda, \beta}^\dagger(t')\}\rangle$. 
We use labels $i$ and $j$ for sites of the system, and $\alpha$ and $\beta$ for sites of the filter. Fourier transforming using ${\bf G}_{\rm S}(t, t') = \sum_m \int \frac{d\omega}{2\pi}\, e^{-i m \Omega t}e^{-i \omega (t - t')}{\bf G}_{\rm S}^{(m)}(\omega)$, where the appearance of the discrete index $m$ follows from the discrete time translation symmetry of the system, we find
\bea
\label{eq:FloquetLandauer}&\bar{I} = \frac{e}{h} \sum_m \int d\omega \left[n_r(\omega) T^{(m)}_{\ell r}(\omega) - n_\ell(\omega) T^{(m)}_{r\ell}(\omega)\right],\ \ \ \\
\nonumber &T^{(m)}_{\lambda\lambda'}(\omega) \equiv {\rm Tr}\left[ {\bf G}_{\rm S}^{(m)\dagger}(\omega){\tilde{\boldsymbol{\Gamma}}_{\rm F,\lambda}(\omega + m \Omega)}{\bf G}_{\rm S}^{(m)}(\omega)\boldsymbol{\Gamma}_{\rm F,\lambda'}(\omega)\right].  
\eea
The energy-filtered lead-system couplings $\boldsymbol{\Gamma}_{\rm F}^{(\lambda)}(\omega)$ and $\tilde{\boldsymbol{\Gamma}}_{\rm F}^{(\lambda)}(\omega)$ are given 
by (suppressing $\omega$ indices for brevity):  
\bea
\nonumber \boldsymbol{\Gamma}_{{\rm F},\lambda}(\omega) = {\bf H}_{{\rm SF},\lambda}\vec{G}_{{\rm F},\lambda}(\omega) \boldsymbol{\Gamma}_{{\rm L},\lambda}(\omega)\vec{G}^\dagger_{{\rm F},\lambda}(\omega) {\bf H}^{\dagger}_{{\rm SF},\lambda}\\
\label{eq:filterGamma} 
{\tilde{\boldsymbol{\Gamma}}_{\rm F,\lambda}(\omega) = {\bf H}_{\rm SF,\lambda}\vec{G}^{\dagger}_{\rm F,\lambda}(\omega) \boldsymbol{\Gamma}_{\rm L,\lambda}(\omega) \vec{G}_{\rm F,\lambda}(\omega){\bf H}^{\dagger}_{\rm SF,\lambda}},
%
\eea
where $\boldsymbol{\Gamma}_{\rm L,\lambda}(\omega) = 2\pi {\bf H}_{\rm FL,\lambda} \boldsymbol{\rho}_\lambda(\omega){\bf H}^{\dagger}_{\rm FL,\lambda}$ is the familiar lead coupling appearing in the standard Landauer formula, with $\boldsymbol{\rho}_\lambda(\omega) = \sum_\nu \Ket{\lambda\nu}\Bra{\lambda\nu}\delta(\omega - \epsilon_{\lambda\nu})$ encoding the density of states of lead $\lambda$. At zero temperature and bias voltage $V$, the differential conductance is given by $\frac{\partial\bar{I}}{dV}\Bigr|_{\substack{V=0}}=\sum_m(T_{\ell r}^{(m)}+T_{r\ell}^{(m)})$.

Equations (\ref{eq:FloquetLandauer}) and (\ref{eq:filterGamma}) along with the ``Floquet distribution function" in Equation (\ref{eq:dist_fn}) below  are the main analytical results of this paper. 
The coefficient $T^{(m)}_{\lambda\lambda'}(\omega)$ describes the probability for an electron emanated from lead $\lambda'$ at energy $\hbar\omega$ to be transmitted across the system to lead $\lambda$ while absorbing $m$ photons from the driving field.
Comparing to the standard Landauer formula, the appearance of $\tilde{\bf \Gamma}_{\rm F,\lambda}(\omega)$ and ${\bf \Gamma}_{\rm F,\lambda'}(\omega)$ in the expression for $T^{(m)}_{\lambda\lambda'}(\omega)$ shows how the net effect of introducing the energy filters is transparently manifested in the transport equation (\ref{eq:FloquetLandauer}) via renormalized effective system-lead coupling matrices.
Due to the narrow bandwidth of the filter, $\vec{G}_{\rm F,\lambda}(\omega)$ and hence $\tilde{\vec{\Gamma}}_{\rm F,\lambda}(\omega)$ are exponentially suppressed for energies outside of the filter band. Therefore, for an ideal filter, all the transmission coefficients in Eq.~(\ref{eq:FloquetLandauer}) are suppressed except for $T_{\lambda\lambda'}^{(0)}$ and the conductance acquires a similar form as in equilibrium.

In Fig.~\ref{fig:filter_sketch}c we illustrate the energy filtering effect of a single filter for the simple case of a one dimensional system. 
Here the filter is modeled as a  nearest-neighbor tight-binding chain of $L_{\rm F}$ sites, with hopping parameter $J$.
The lead is connected to the first site of the chain ($\alpha = 1$), while the system is coupled to the last site of the chain ($\alpha = L_{\rm F}$).
We consider the lead in the wide-band limit~\cite{Kohler2005}, which gives a frequency-independent coupling to the first site: $[\boldsymbol{\Gamma}_{\rm L}(\omega)]_{\alpha \beta} = \gamma\delta_{\alpha 1}\delta_{\beta 1}$.
Under these conditions, the energy dependence of ${\bf \Gamma}_{\rm F}(\omega)$ is controlled by $[{\bf G}_{\rm F}(\omega)]_{L_{\rm F}1}$ --- the Green's function component that connects the two ends of the filter.
In Fig.~\ref{fig:filter_sketch}c we plot $\log_{10} |[{\bf G}_{\rm F}(\omega)]_{L_{\rm F}1}|$ as a function of $\omega$ and filter length, $L_{\rm F}$.
Outside the filter bandwidth, $|[{\bf G}_{\rm F}(\omega)]_{L_{\rm F}1}|$ drops rapidly (with an amplitude that falls exponentially with the length of the filter); for energies within the filter window, the transmission is close to 1.
These are precisely the features that we expect can help to restore nearly quantized transport through Floquet topological edge modes, see above and Fig.~\ref{fig:filter_sketch}b.

\begin{figure}[t]
\includegraphics[width=\columnwidth]{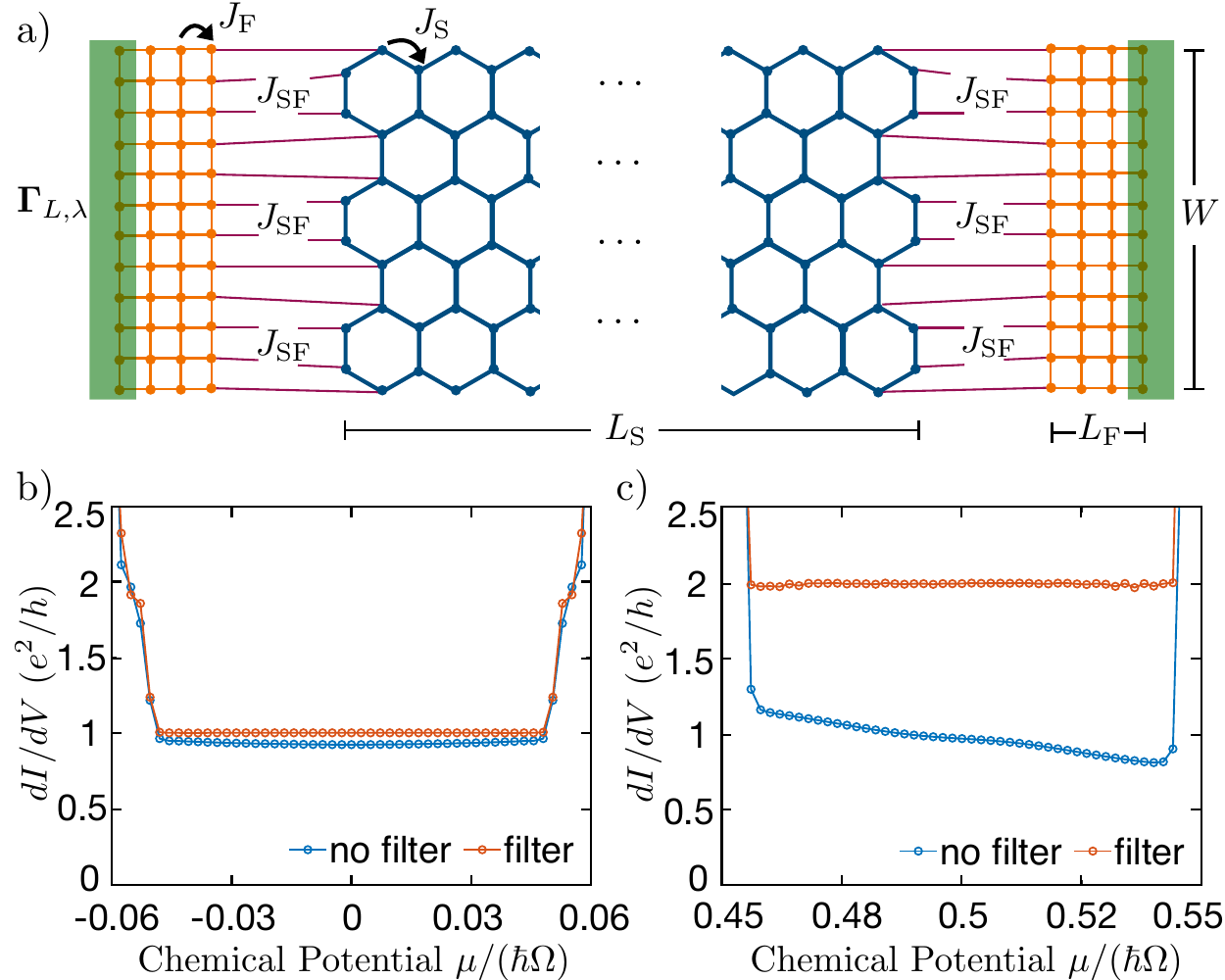}
	\caption{ a) Sketch of the system setup.  
    We set the system hopping to $J_{\rm S}=1$; all other energies are scaled relative to $J_{\rm S}$. 
    Throughout we set the drive amplitude to $A_0=0.5$, its frequency to $\hbar\Omega=3.25$, and the filter hopping amplitude to $J_{\rm F}=0.25$.
    The latter ensures that the filter bandwidth is less than $\hbar\Omega$. 
    The system-filter coupling $J_{\rm SF}$ is set to the geometric mean of $J_{\rm S}$ and $J_{\rm F}$. The filter-lead coupling is set to $\gamma=0.25$.
    b), c) Numerical simulations of the differential conductance with $W=160$ sites, $L_{\rm S}=120$ sites and $L_{\rm F}=20$ sites in the off-resonant (b) and resonant (c) gaps.  
	}
\label{fig:results}
\end{figure}

{\it Numerical simulations.}--- 
We numerically investigate transport through a graphene-like nearest neighbor tight-binding system of width $W$ and length $L_{\rm S}$, with zig-zag edges in the direction from one lead to the other (see Fig.~\ref{fig:results}a).
A circularly polarized vector potential $\vec{A}(t) = A_0(\cos(\Omega t), \sin(\Omega t), 0)$ acting on the system introduces time-dependent Peierls' phases to the hopping matrix elements: 
$H_{\rm S}(t) = J_{\rm S}\sum_{\vec{r}_i}\sum_{\sigma}e^{-ie/\hbar\vec{A}(t)\cdot \vec{b}_\sigma} 
 c^{\dagger}_{\vec{r}_i+\Vec{b}_\sigma}c_{\vec{r}_i} + h.c. + H_{\rm BG}$, where $\vec{r}_i$ runs over all the lattice sites and $\vec{b}_\sigma$ runs over the three nearest hopping vectors, $J_{\rm S}$ is the nearest hopping amplitude and $H_{\rm BG}$ models a backgate voltage 
 as a uniform potential on all sites.
 We use single band square lattices to model the energy filters, each with width $W$ equal to that of the system and length $L_{\rm F}$. 
 We fix the lead chemical potential at the center of the filter band and vary the backgate voltage on the system to probe transport over a range of energies. 
 
The leads are connected to the first (last) slice of the left (right) filter, as indicated via green shading in Fig.~\ref{fig:results}a. Taking the wide-band limit for the lead, the coupling matrix $\vec{\Gamma}_{\rm L,\lambda}(\omega)$ takes a similar form as in the one dimensional case, now with a single parameter $\gamma$ appearing on all diagonal matrix elements corresponding to sites in the shaded slice (with all other matrix elements zero). 
For simulations without the filters, the leads connect to the system directly; here $\vec{\Gamma}_{\rm F,\lambda}$ and $\tilde{\vec{\Gamma}}_{\rm F,\lambda}$ in Eq.(\ref{eq:FloquetLandauer}) are replaced by $\vec{\Gamma}_{\rm L,\lambda}$. 
 We compute the relevant Green's functions using the recursive Green's function algorithm~\cite{RecurGreen}; see Appendix \ref{app:FloquetGreen} for details of the numerical implementation. 
 
The differential conductance in the absence of filters is shown as the blue curves in Figs.~\ref{fig:results} b) and c), corresponding to transport through the off-resonant and resonance gaps, respectively. Note that the resonance gap hosts {\it two} chiral modes (see Fig.~\ref{fig:summary}a) and thus the differential conductance is naively expected to be quantized to $2e^2/h$.
Even though the drive induces chiral edge modes, the differential conductance is far from quantized in both cases due to photon assisted transport~\cite{Farrell2015,Farrell2016}. 
 
Results for transport with the energy filters are shown as the orange curves in Figs.~\ref{fig:results} b,c. As designed, the filters restore the differential conductance in the off-resonant (resonance) gap to a quantized value of $e^2/h$ ($2e^2/h$).

 
\begin{figure}[t]
	\includegraphics[width=\columnwidth]{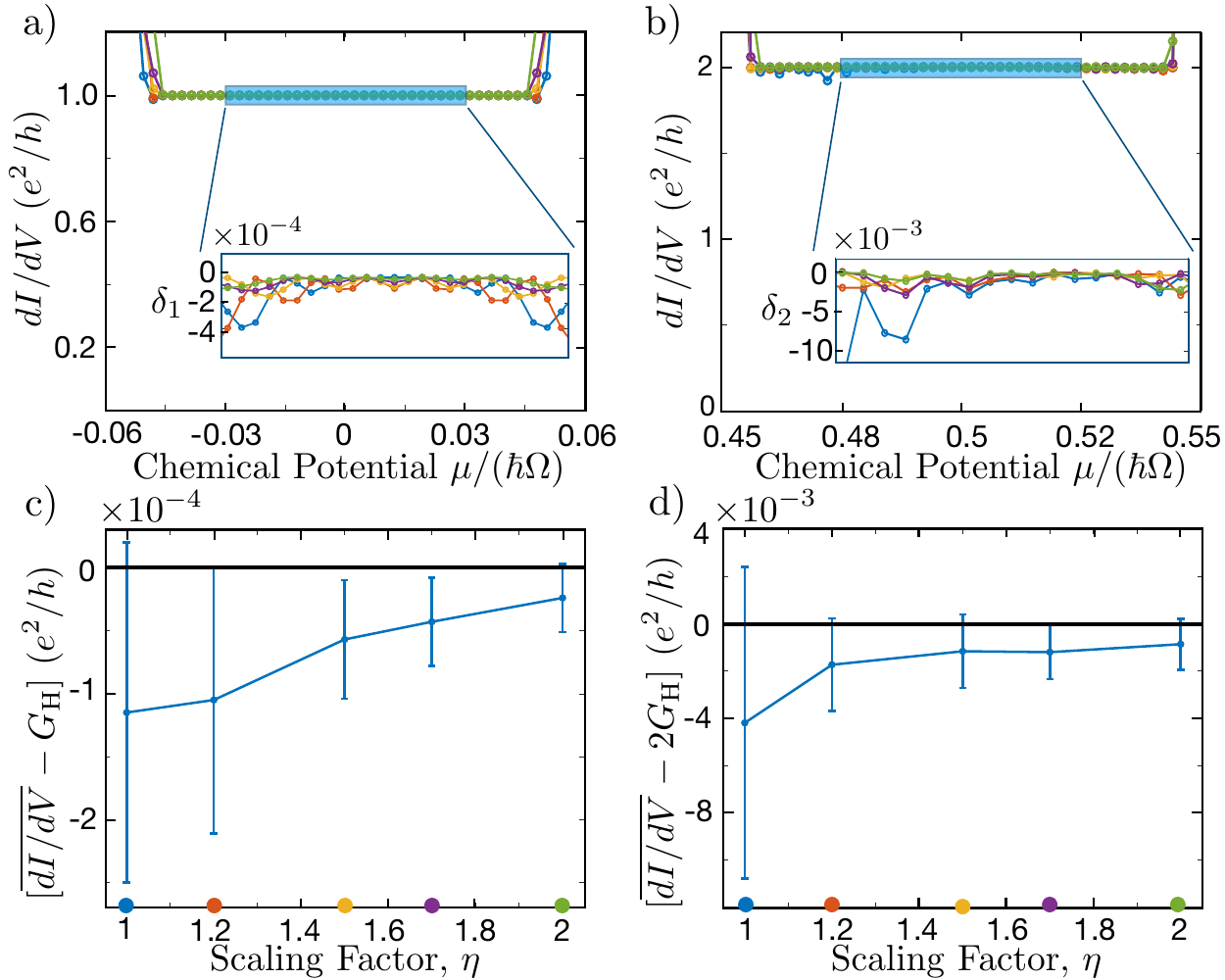}
	\caption{Characterization of energy-filtered topological transport. 
 We set the initial geometry to be $L_{\rm S}\times W=120\times120$ and $L_{\rm F}=10$. 
 Fixing $L_S$, we scale up $W$ and $L_{\rm F}$ by a factor $\eta$.
 All the other parameters are the same as in earlier figures.  
 Differential conductance vs.~scaling factor in the off-resonant (a) and resonance (b) gaps. 
 Insets: zoom-ins of the blue shaded regions; $\delta_1$ and $\delta_2$ are the deviations from $e^2/h$ and $2e^2/h$, respectively. 
 Panels c) and d) show the average differential conductance and its variance in the off-resonant and resonance gaps.
 The colored dots on the horizontal axis mark the scaling factors for the curves in a) and b). 
}
\label{fig:filter_performance}
\end{figure}

 We now characterize the quality of quantization and its dependence on the filter's geometry.
We start with an $L_{\rm S}\times W=120 \times 120$ system and a filter of $L_{\rm F}=10$ sites. 
We then compute the differential conductance while scaling up the width $W$ and the filter length $L_{\rm F}$ by a scale factor $\eta$, while fixing the system length at $L_{\rm S} = 120$ sites [Figs.~\ref{fig:filter_performance}a,b]. 
To quantify the degree of quantization, we calculate the average differential conductance across the plateaus within the off-resonant and resonance gaps, and the associated variances.
As shown in Figs.~\ref{fig:filter_performance}c,d, 
the differential conductance in the off-resonant (resonance) gap converges toward $e^2/h$ ($2e^2/h$) with increasing $\eta$; the average and variance (shown by the error bars) are 
taken over the blue shaded regions in panels a) and b). 
For system dimensions of order 100 sites, the filters improve quantization by approximately three orders of magnitude compared to the case without filters [Figs.~\ref{fig:results}b,c].

\begin{figure}[t]
	\includegraphics[width=\columnwidth]{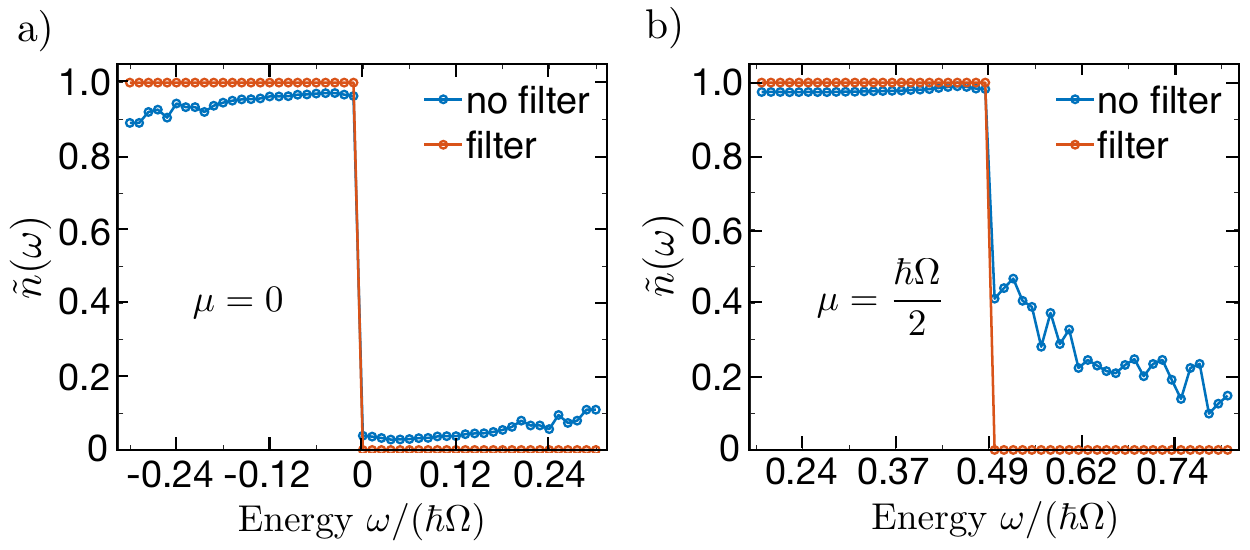}
	\caption{The non-equilibrium ``distribution function" $\Tilde{n}(\omega)$, Eq.~(\ref{eq:FloquetDist}), for the filtered and non-filtered systems. 
 With the filter in place, the distribution $\Tilde{n}(\omega)$ closely approximates a Fermi function, coinciding with the restoration of quantized transport. a) Distribution function for $\mu = 0$ (for transport in the off-resonant gap). b) Distribution function for $\mu = \hbar\Omega/2$ (for transport in the resonant gap). 
}
\label{fig:filter_system_coupling}
\end{figure}

{\it Distribution function.}--- In the non-driven case, quantized transport follows from the Fermi distributions in the chiral edge modes. 
In equilibrium, the distribution function can be obtained from the relation between the lesser and the retarded Green's function \cite{Bruus2004}: ${\rm Im}[G^{<}(\omega)] = -2{\rm Im}[G(\omega)] n(\omega)$. 
In driven systems however, 
we lack such a simple and general relation. 
Inspired by the equilibrium case, we define a non-equilibrium distribution function in terms of the time-averaged retarded and lesser Green's functions, $\vec{G}_{\rm S}^{(0)}(\omega)$ (see above) and $\vec{G}_{\rm S}^{<(0)}(\omega)=\frac{1}{T}\int_0^Tdt\int dt' e^{i\omega(t-t')}\vec{G}_{\rm S}^{<}(t,t')$, with $\vec{G}_{\rm S}^{<}(t, t')=-i\langle c^{\dagger}(t')c(t) \rangle$: 
\bea
\label{eq:FloquetDist}
\Tilde{n}(\omega) = \frac{{\rm Tr}\big\{{\rm Im}[\vec{G}_{\rm S}^{<(0)}(\omega)]\big\}}{{\rm Tr}\big\{{\rm Im}[\vec{G}_{\rm S}^{(0)}(\omega)]\big\}}. 
\eea
Utilizing the Heisenberg equations of motions for $\vec{a}_\lambda(t)$, $\vec{f}_\lambda(t)$, and $\vec{c}(t)$ (see Appendix A) and taking the time average over one period,  $\Tilde{n}(\omega)$ can be expressed as
\bea
\label{eq:dist_fn}
\Tilde{n}(\omega) = -\frac{\sum_{\lambda, m}{\rm Tr}\big\{{\rm Im}[\vec{D}_\lambda^{(m)}\vec{\Gamma}_{\rm L,\lambda}\vec{D}_\lambda^{(m)\dagger}n_{\lambda}]\big\}(\omega-m\Omega)}
{2\sum_{\lambda, m'}{\rm Tr}\big\{{\rm Im}[\vec{D}_\lambda^{(m')}\vec{\Gamma}_{\rm L,\lambda}\vec{D}_\lambda^{(m')\dagger}]\big\}(\omega-m'\Omega)},\ \ \ \
\eea
where 
$\vec{D}_{\lambda}^{(m)}$ is given by 
$\vec{D}_{\lambda}^{(m)}=\vec{G}_{\rm S}^{(m)}\vec{H}_{\rm SF,\lambda}\vec{G}_{\rm F,\lambda}$. By definition, $\vec{\Gamma}_{\rm L,\lambda}$ is always a positive semi-definite matrix. 
Thus $\vec{D}_\lambda^{(m)}\vec{\Gamma}_{\rm L,\lambda}\vec{D}_\lambda^{(m)\dagger}$ can be decomposed into a matrix multiplied by its own Hermitian conjugate. 
As a result, the trace is always positive and we necessarily have $0\le\Tilde{n}(\omega)\le1$ as desired for a distribution function. 

Without the filter, the behaviour of $\Tilde{n}(\omega)$ results from a complicated interplay of contributions from all harmonics. 
When the filter is in effect, however, all the $m,m'\neq0$ terms in Eq.~(\ref{eq:dist_fn}) will have frequencies outside the filter band 
and will thus be exponentially suppressed. 
The surviving $m=0$ term gives the Fermi distribution $n(\omega)$ (for $\mu_{\ell}=\mu_r$). The non-equilibrium distribution function $\Tilde{n}(\omega)$ in the off-resonant and resonance gaps is shown in Fig.~\ref{fig:filter_system_coupling}. 
As Eq.~(\ref{eq:dist_fn}) predicted, a Fermi distribution is restored when the filters are inserted. 
These results for the non-equilibrium distribution function highlight the mechanism that restores quantization.

{\it Discussion.}--- In conclusion, narrow band filters inserted between the leads and system effectively suppress  photon assisted transport in Floquet systems.
For topological Floquet systems with chiral or helical edge modes, this phenomenon can be used to elicit quantized transport from the non-equilibrium system. 
Our numerical studies focused on the regime of a single resonance within the bands. 
When multiple resonances are present, energy filters may still be used to help establish a sharp Fermi distribution across all chiral edge states within a given quasi-energy gap. Further exploration of this regime is an interesting direction for future work.
As demonstrated by our numerics, the precision of the quantization improves as the filter is enlarged. We expect that energy filters present a promising direction for experiments in the emerging area of transport in cold atomic gases in driven optical lattices.
A narrow impurity band or a moir\'{e} flat band in solid systems could also be suitable candidates for the energy filters.

{\it Acknowledgements.}--- We thank Mads Kruse for helpful discussions and preliminary contributions in the development of this work.
M. R. gratefully acknowledges the support of the European Research Council (ERC) under the
European Union Horizon 2020 Research and Innovation Programme (Grant Agreement No. 678862) and the Villum Foundation, as well as the Brown Investigator Award, a program of
the Brown Science Foundation, the University of Washington College of Arts and Sciences, and the Kenneth K.
Young Memorial Professorship.
N. L. is grateful for funding from the ISF Quantum Science and Technology program (2074/19).

\bibliography{References}

\clearpage
\begin{appendix}
\begin{widetext}
    

\section{Derivation of the filtered Floquet Landauer formula}\label{app:Landauer}

In this section we give the detailed derivation for the Floquet Landauer formula in Eq. \eqref{eq:FloquetLandauer}, following an analogous approach to that used for example in Sec.~3 of Ref.~\onlinecite{Kohler2005}.
We begin by writing the expression for the time-averaged current that flows between the filter and the system:

\be
\label{eq: average current}
\bar{I} =  \frac{1}{T} \int_0^T dt\, \frac{2e}{\hbar} {\rm Im} \avg{{\bf c}^\dagger(t){\bf H}_{{\rm SF},r} {\bf f}_r(t)}.
\ee

To evaluate 
Eq.~\eqref{eq: average current}, we solve the Heisenberg equations of motion for the operators ${\bf a}_\lambda(t)$, ${\bf f}_\lambda(t)$ and ${\bf c}(t)$: 
\bea
\label{eq:adot1} 
i\hbar \dot{\vec{a}}_\lambda &=& {\bf H}_{{\rm L},\lambda} \vec{a}_{\lambda} + {\bf H}^{\dagger}_{\rm FL,\lambda}\vec{f}_\lambda,\\
\label{eq:fdot1}
i\hbar \dot{\bf f}_{\lambda} &=& {\bf H}_{\rm F,\lambda} {\bf f}_{\lambda} + {\bf H}_{\rm FL,\lambda} {\bf a}_{\lambda} + {\bf H}^{\dagger}_{\rm SF,\lambda} {\bf c},\\
\label{eq:cdot1} 
i\hbar \dot{\bf c} &=& {\bf H}_{\rm S} {\bf c} + \sum_{\lambda} {\bf H}_{\rm SF,\lambda} {\bf f}_\lambda.
\eea
Our goal is to ultimately express the expectation values of the system and filter operators in Eq.~(\ref{eq: average current}) in terms of expectation values of operators in the leads in the distant past.
By assuming each lead was in thermal equilibrium in the distant past, we will then be able to evaluate the current via the expectation value in Eq.~(\ref{eq: average current}), see below.

We employ the Green's function method to solve Eqs.~(\ref{eq:adot1}) - (\ref{eq:cdot1}). 
 We define the ``bare'' Green's function of lead $\lambda$, $\vec{g}_\lambda(t-t')$, via
\begin{equation}
        \label{eq:g_lambda}
        \left[i\frac{\partial}{\partial t}-\vec{H}_{\rm L,\lambda}\right]\vec{g}_\lambda(t-t')=\mathbf{1}_{\rm L}\delta(t-t');\  \quad \vec{g}_\lambda(t-t')=0 \ {\rm for}\ t<t'. 
\end{equation}  
The solution to Eq.~(\ref{eq:g_lambda}) is $g_\lambda(t - t') = -i\theta(t-t')e^{-iH_{{\rm L},\lambda}(t-t')}$.
Using Eq.~(\ref{eq:g_lambda}), we formally integrate Eq.~(\ref{eq:adot1}) to obtain
\begin{equation}
        \label{eq:a_soln}
        \vec{a}_\lambda(t) = e^{-i {\bf H}_{\rm L,\lambda} (t-t_0)}\vec{a}_\lambda(t_0) + \int_{-\infty}^\infty dt' \vec{g}_\lambda(t - t') {\bf H}^{\dagger}_{\rm FL,\lambda} \vec{f}_\lambda(t').
\end{equation}

We furthermore define the {\it full} Green's function of filter $\lambda$, 
$\vec{G}_{\rm F,\lambda}(t-t')$, 
which satisfies
\begin{equation}
    \begin{aligned}
    \label{eq:f_soln}
        &\left[i\frac{\partial}{\partial t}-\vec{H}_{{\rm F},\lambda}\right]\vec{G}_{\rm F,\lambda}(t-t')-\int_{-\infty}^{\infty}dt''\vec{\Sigma}_{\rm FLF,\lambda}(t-t'')\vec{G}_{\rm F,\lambda}(t''-t') = \mathbf{1}_{\rm F}\delta(t-t'), \\
        &\vec{f}_\lambda(t) = 
        \int_{-\infty}^\infty dt' \vec{G}_{\rm F,\lambda}(t - t') [\vec{h}_\lambda(t')+{\bf H}_{\rm SF,\lambda}^{\dagger}\vec{c}(t')],
    \end{aligned}
\end{equation}
where $\vec{\Sigma}_{\rm FLF,\lambda}(t-t'')=\vec{H}_{\rm FL,\lambda}\vec{g}_\lambda(t-t'')\vec{H}_{\rm FL,\lambda}^{\dagger}$ is the self-energy of the filter and $\vec{h}_\lambda(t') = \vec{H}_{\rm FL,\lambda}e^{-i\vec{H}_{\rm L,\lambda}(t'-t_0)}\vec{a}_\lambda(t_0)$. 
Similarly, the full Green's function of the system, $\vec{G}_{\rm S}(t,t')$, satisfies
\begin{equation}
    \begin{aligned}
    \label{eq:c_soln}
        &\left[i\frac{\partial}{\partial t}-\vec{H}_{\rm S}(t)\right]\vec{G}_{\rm S}(t,t')-\sum_\lambda\int_{-\infty}^{\infty}dt''\vec{\Sigma}_{\rm SFS,\lambda}(t-t'')\vec{G}_{\rm S}(t'', t')=\mathbf{1}_{\rm S}\delta(t-t'),\\
        &\vec{c}(t) = 
        \sum_\lambda\int_{-\infty}^\infty dt' \int_{-\infty}^\infty dt'' \vec{G}_{\rm S}(t,t'){\bf H}_{\rm SF,\lambda}\vec{G}_{\rm F,\lambda}(t'-t'')\vec{h}_\lambda(t'').
    \end{aligned}
\end{equation}
Here  $\vec{\Sigma}_{\rm SFS,\lambda}(t-t'')=\vec{H}_{\rm SF,\lambda}\vec{G}_{\rm F,\lambda}(t-t'')\vec{H}_{\rm SF,\lambda}^{\dagger}$ is the self-energy of the system. 

In writing Eqs.~(\ref{eq:f_soln}) and (\ref{eq:c_soln}), we have assumed that the initial time $t_0$ is in the distant past.
Due to the finite decay times of the Green's functions and self energies appearing inside the integrals, we have therefore extended the lower limits of integration to $-\infty$.
Moreover, we have assumed that on this timescale any memory of the initial conditions within the system or the filter has been erased: $\vec{G}_{\rm F,\lambda}(t-t_0)\vec{f}_\lambda(t_0) \to 0$ and  $\vec{G}_{\rm S}(t,t_0)\vec{c}(t_0) \to 0$.

Before proceeding, we note that the filter Green's function $\vec{G}_{\rm F,\lambda}$ obtained from the equations of motion and the one defined in the main text are equivalent. To show this, differentiate $[\vec{G}_{\rm F,\lambda}(t, t')]_{\alpha \beta}=-i\theta(t-t')\langle \{{f}_{\lambda, \alpha}(t), {f}_{\lambda, \beta}^\dagger(t')\}\rangle$ with respect to $t$. Utilizing Eqs.~(\ref{eq:adot1}), (\ref{eq:fdot1}), (\ref{eq:cdot1}), it can be shown that the results exactly match the first line of Eq.~(\ref{eq:f_soln}). Similarly, the system Green's function $\vec{G}_{\rm S}$ defined here is equivalent to the one defined in the main text.

\renewcommand\thesubsection{\thesection.\arabic{subsection}}
\subsection{Evaluating expectation values of the source terms $\vec{h}_\lambda(t)$ and $\vec{h}^\dagger_\lambda(t)$}

The operator $\vec{h}_\lambda(t)$ defined below Eq.~(\ref{eq:f_soln}) acts as a source term for the operators ${\bf f}(t)$ and ${\bf c}(t)$ in Eqs.~(\ref{eq:f_soln}) and (\ref{eq:c_soln}), carrying information about the initial state of lead $\lambda$ to the filter and eventually to the system. 
Using Eqs.~(\ref{eq:f_soln}) and (\ref{eq:c_soln}), we will express the current in Eq.~(\ref{eq: average current}) 
in terms of $\vec{h}_\lambda^{\dagger}$ and $\vec{h}_\lambda$. 
Thus, to facilitate the evaluation of the current, we will first study the expectation values of products of components of $\vec{h}_\lambda^{\dagger}$ and $\vec{h}_\lambda$.
 For later use, it will be most convenient to work with the Fourier transformed operators $\vec{h}_\lambda(\omega)=\int_{-\infty}^{\infty} dt \, e^{i\omega t}\vec{h}_\lambda(t)$. 


Consider a generic expectation value $\avg{h_{\lambda,m}^{\dagger}(\omega)h_{\lambda',n}(\omega')}$, where $m$ ($n$) labels states in an arbitrary orthonormal basis for filter $\lambda$ ($\lambda'$).
Using the definition of $\vec{h}_\lambda$ and expanding (without loss of generality) in the basis of energy eigenstates $\{\Ket{\lambda \nu}\}$ of each lead, ${\bf H}_{\rm L,\lambda}\Ket{\lambda \nu} = \varepsilon_{\lambda\nu}\Ket{\lambda \nu}$,  we obtain: 
\begin{equation}
    \begin{aligned}
        \avg{h_{\lambda,m}^{\dagger}(\omega)h_{\lambda',n}(\omega')} &=     
        \sum_\nu \sum_{\nu'} \int dt\, e^{-i\omega t} e^{i\varepsilon_{\lambda\nu}t} \int dt'e^{i\omega't'}e^{-i\varepsilon_{\lambda'\nu'}t'}[{\bf H}_{\rm FL,\lambda}^{\dagger}]_{\nu m} [{\bf H}_{\rm FL,\lambda'}]_{n\nu'} \avg{a_{\lambda\nu}^{\dagger}(t_0)a_{\lambda'\nu'}(t_0)}\\
        &=(2\pi)^2\sum_\nu\delta_{\lambda\lambda'}\delta(\omega-\omega')\delta(\omega-\varepsilon_{\lambda\nu})[{\bf H}_{\rm FL,\lambda}^{\dagger}]_{\nu m}[{\bf H}_{\rm FL,\lambda}]_{n\nu}n_{\lambda}(\varepsilon_{\lambda\nu}).
    \end{aligned} 
    \label{eq:hexpec}
\end{equation}
To arrive at the second line, we used the assumption that each lead was in equilibrium in the distant past, captured by $\langle a_{\lambda\nu}^\dagger(t_0) a_{\lambda'\nu'}(t_0)\rangle=\delta_{\lambda\lambda'}\delta_{\nu\nu'} n_{\lambda}(\epsilon_{\lambda\nu})$, where $n_{\lambda}$ is the Fermi distribution for lead $\lambda$.

Next, it is helpful to introduce the effective lead coupling $\vec{\Gamma}_{\rm L,\lambda}(\omega)=2\pi\vec{H}_{\rm FL,\lambda}\vec{\rho}_\lambda(\omega)\vec{H}_{\rm FL,\lambda}^{\dagger}$, and the operator $\vec{\rho}(\omega)=\sum_\nu\Ket{\lambda\nu}\Bra{\lambda\nu}\delta(\omega-\varepsilon_{\lambda\nu})$. 
For any matrix $\vec{A}$ acting on the Hilbert space of filter $\lambda$, we have
\be
\label{eq:h_exp_gen}\avg{\vec{h}_\lambda^\dagger(\omega)\vec{A}\vec{h}_{\lambda}(\omega')} = 2\pi \delta(\omega-\omega'){\rm Tr}[\vec{A}\vec{\Gamma}_{\rm L,\lambda}(\omega)]n_{\lambda}(\omega).
\ee
In the next subsection, we will use this general expression to evaluate the expectation value of the current, Eq.~(\ref{eq: average current}).

\subsection{Evaluation of the current}
In Eqs.~(\ref{eq:a_soln})-(\ref{eq:c_soln}) we obtained a set of closed expressions for the Heisenberg operators $\vec{a}_\lambda(t)$, $\vec{f}_\lambda(t)$, and $\vec{c}(t)$. 
Substituting the expression for $\vec{f}_{\lambda}(t)$ in Eq.~(\ref{eq:f_soln}) into Eq.~(\ref{eq: average current}) gives 
\bea
    \begin{aligned}
        \label{eq:I_subbed} \bar{I} = \bar{I}_1+\bar{I}_2 =\frac{1}{T}\frac{2e}{\hbar}\int_{0}^{T}dt\int_{-\infty}^{\infty} dt'\, \left[ {\rm Im}\avg{\vec{h}_{r}^{\dagger}(t)\vec{G}_{{\rm F},r}^{\dagger}(t-t')\vec{H}_{{\rm SF},r}^{\dagger} \vec{c}(t')} +{\rm Im}\avg{\vec{c}^\dagger(t)\vec{\Sigma}_{{\rm SFS},r}^{\dagger}(t-t')\vec{c}(t')}\right].
    \end{aligned}
\eea
The steps for evaluating the two integrals $I_1$ and $I_2$ in Eq.~(\ref{eq:I_subbed}) are similar; 
here we only give the detailed procedure for the first one. 
The next step is to substitute the expression for $\vec{c}(t)$ in Eq.~(\ref{eq:c_soln}) into the Eq.~(\ref{eq:I_subbed}).
After Fourier transformation, the integrand takes the form of the general expectation value in Eq.~(\ref{eq:h_exp_gen}).
%
%
%
Using Eq.~(\ref{eq:h_exp_gen}), we find 
\be
    \label{eq:eval_first}
    \bar{I}_1 = \frac{e}{h}\frac{1}{T}\int_0^{T }dt \int_{-\infty}^{\infty} d\omega \,{\rm Im}\biggl\{{\rm Tr}\left[ \vec{G}_{{\rm F},r}^{\dagger}(\omega) \vec{H}_{{\rm SF},r}^{\dagger} \vec{G}_{\rm S}(t,\omega) \vec{H}_{{\rm SF},r}\vec{G}_{{\rm F},r}\vec{\Gamma}_{{\rm L},r}(\omega)\right]\biggr\}n_{r}(\omega).
\ee
Here we have used the Fourier transform $\vec{G}_{\rm S}(t,\omega)=\int dt'e^{i\omega(t-t')}\vec{G}_{\rm S}(t, t')$. 

Using the fact that $\vec{\Gamma}_{\rm L,\lambda}$ is Hermitian by definition, taking the imaginary part of the trace in Eq.~(\ref{eq:eval_first}) brings out a piece $[\vec{G}_{\rm S}(t,\omega)-\vec{G}_{\rm S}^{\dagger}(t,\omega)]/2i$. 
To evaluate this we need to utilize the first line of Eq.~(\ref{eq:c_soln}). 
The trick is the following: first use the Fourier transform defined above on both sides to obtain an equation for $\vec{G}_{\rm S}(t,\omega)$; multiply $\vec{G}_{\rm S}^{\dagger}(t,\omega)$ from the left to get the first temporary expression; complex conjugate the first line of Eq.~(\ref{eq:c_soln}) and then multiply $\vec{G}_{\rm S}(t,\omega)$ from the right to get the second temporary expression.
Subtract these two expressions, then expand $\vec{G}_{\rm S}(t,\omega)=\sum_m e^{-im\Omega t} \vec{G}_{\rm S}^{(m)}(\omega)$ in a discrete Fourier series [noting the time-periodicity of ${\bf G}_{\rm S}(t, \omega)$], and finally take the time average over one period: 
\be
    \begin{aligned}
        \frac{1}{T}\int_0^T dt \left[\, \vec{G}_{\rm S}(t,\omega)-\vec{G}_{\rm S}^{\dagger}(t,\omega)\right] &= \sum_\lambda \sum_m \vec{G}_{\rm S}^{(m)\dagger}(\omega)[\vec{\Sigma}_{\rm SFS,\lambda}-\vec{\Sigma}_{\rm SFS,\lambda}^{\dagger}](\omega+m\Omega)\vec{G}_{\rm S}^m(\omega)\\
        &= \sum_\lambda \sum_m \vec{G}_{\rm S}^{(m)\dagger}(\omega) \vec{H}_{\rm SF,\lambda} [\vec{G}_{\rm F,\lambda}-\vec{G}_{\rm F,\lambda}^{\dagger}](\omega+m\Omega) \vec{H}_{\rm SF,\lambda}^{\dagger} \vec{G}_{\rm S}^{(m)}(\omega).
        \label{eq:Imag_trick}
    \end{aligned}
\ee
Applying the same trick on $\vec{G}_{F,\lambda}-\vec{G}_{F,\lambda}^{\dagger}$ in Eq.~(\ref{eq:Imag_trick}), but using the first line of Eq. (\ref{eq:f_soln}), we arrive at our final expression: 
\be
    \bar{I}_1 = \frac{e}{h} \sum_\lambda \sum_m \int_{-\infty}^{\infty} d\omega \,{\rm Tr}[\vec{G}_{\rm S}^{(m)\dagger} \tilde{\vec{\Gamma}}_{\rm F,\lambda}(\omega+m\Omega) \vec{G}_{\rm S}^{(m)}(\omega) \vec{\Gamma}_{{\rm F},r}(\omega)]n_{r}(\omega),
\ee
where  $\vec{\Gamma}_{\rm F}(\omega)$ and  $\tilde{\vec{\Gamma}}_{\rm F}(\omega)$ are as defined in the main text.
After a similar procedure on the second term of Eq.~(\ref{eq:I_subbed}) and adding up the two terms, we obtain the Floquet Landauer formula given in Eq.~(\ref{eq:FloquetLandauer}) of the main text. 

\end{widetext}

\section{Numerical calculation of the Floquet Green's function}\label{app:FloquetGreen}
In this section we discuss the detailed implementation of our numerical simulations. 
The main difficulties in this are time dependence of the Hamiltonian and inverting large matrices of size many times larger than the Hilbert space dimension. 
Below we review the 
recursive Green's function algorithm {\cite{RecurGreen}} and discuss its adaptation to the extended space representation for Floquet systems~\cite{Oka2019, FloquetHandbook} to address these two challenges. 

The recursive Green's function algorithm is commonly used to calculate transport properties in equilibrium systems with open boundary conditions. 
The algorithm breaks the lattice into $N$ slices and recursively computes the Green's functions for each individual slice.
Components of the full Green's function can then be obtained by combining the slice Green's functions in appropriate ways. 
This algorithm avoids inverting large matrices by performing many inversions in the subspace of individual slices, thus speeding up computations dramatically. 

The recursive Green's function algorithm was developed for equilibrium (non-driven) systems.
To implement it for Floquet systems, we  utilize the extended space representation (see, e.g., Refs.~\cite{Sambe1972, FloquetHandbook}). Importantly, this representation reformulates the time dependent Schrodinger equation into an effective time independent one.
The equation for the full Floquet Green's function can be written in a compact and convenient form (see, e.g., Ref.~\onlinecite{Kitagawa2011}):
\be
[\omega + \vec{\Omega} - \vec{H} - \vec{\Sigma}(\omega)]{\boldsymbol{\mathcal{G}}}(\omega) = \mathbf{1},
\label{eq:floqGreen}
\ee
where 
\be
\boldsymbol{\mathcal{G}}(\omega)=\begin{pmatrix}
    \ddots \\
    & \vec{G}^{0}(\omega-\Omega)&\vec{G}^{-1}(\omega)&\vec{G}^{-2}(\omega+\Omega) \\
    & \vec{G}^{1}(\omega-\Omega)&\vec{G}^{0}(\omega)&\vec{G}^{-1}(\omega+\Omega) \\
    & \vec{G}^{2}(\omega-\Omega)&\vec{G}^{1}(\omega)&\vec{G}^{0}(\omega+\Omega) \\
    & & & & \ddots
\end{pmatrix}.
\label{eq:Gmatrix}
\ee
 Each Fourier component of the Green's function, $\vec{G}^{m}$, is a matrix of the same dimension as the Hilbert space of the system. 
 The matrix $\vec{\Omega}$ has the same block structure as $\boldsymbol{\mathcal{G}}(\omega)$ above, and is zero in all off-diagonal blocks.
 Within diagonal block $n$ [counted in increasing order from the upper left to the lower right, with $n = 0$ corresponding to the block with $\vec{G}^0(\omega)$ in Eq.~(\ref{eq:Gmatrix})], $\vec{\Omega}$ is given by $n\Omega$ times the identity (of dimension equal to that of the system's Hilbert space).


Note that the extended space Green's function $\boldsymbol{\mathcal{G}}(\omega)$ incorporates an infinite redundancy: every column individually carries complete information about the physical Green's function of the system. 
To obtain physical quantities/observables, the physical Green's function components must be extracted from the appropriate blocks of the extended space Green's function.
In particular, to generate the differential conductance, we take the central column of $\boldsymbol{\mathcal{G}}(\omega)$ and combine these Fourier components according to Eq.~(\ref{eq:FloquetLandauer}) of the main text.  

\subsection{Implementation}
We point out that there are two equivalent ways of formulating the transport with filters: a) separate the filter and system Green's functions as in Eq.~(\ref{eq:FloquetLandauer}) and connect them with a system-filter coupling matrix; b) treat everything in between the two leads as one whole system with different local Hamiltonians and coupling matrices for slices in the filter and the system.
The former is useful for highlighting the role of the filter and exposing the physical mechanism through which it helps enable quantized transport in periodically driven systems; 
the latter is more numerically convenient and is implemented in our simulation.

Finally, we give the specific details of our implementation, in particular specifying the form of the lead-system coupling matrix used in our simulations. 
The whole setup (including both the filter and system) is divided into $N=2L_{\rm F}+L_{\rm S}$ vertical slices.
To label a site, we need to specify one index for the slice and another index for the position in that slice. Thus we work in a real space basis that can be written as $\Ket{i}_{\rm H}\otimes \Ket{a}_{\rm V}$, where $i$ labels the horizontal (${\rm H}$) position of a slice and $a$ labels the vertical (${\rm V}$) position of a site in that slice. We label the first slice (indicated by the left green shading in Fig.~\ref{fig:results} of the main text) as $i=1$ and the last slice (right green shading in Fig.~\ref{fig:results}) as $i=N$.
We have assumed that the system is coupled to the lead only via the first and last slices, and that the imaginary part of the self-energy takes the form $\Bra{ia}\vec{\Gamma}_1\Ket{jb}=\gamma\delta_{iN}\delta_{jN}\delta_{ab}$ and $\Bra{ia}\vec{\Gamma}_2\Ket{jb}=\gamma\delta_{i1}\delta_{j1}\delta_{ab}$. The coupling matrices $\vec{\Gamma}_1$($\vec{\Gamma}_2$) take the form $\gamma\mathbf{I}_{\rm V}$ in the block that corresponds to the first (last) slice ($i=j=1$ or $i=j=N$), where $\mathbf{I}_{\rm V}$ is the identity matrix in the subspace of that slice; all other elements are zero. As a result, the trace in Eq.~(\ref{eq:FloquetLandauer}) of the main text is reduced to two partial traces and further simplifies to:
\be
    \begin{aligned}
        {\rm Tr}[\vec{G}^{(m)\dagger}\vec{\Gamma}_1\vec{G}^{(m)}\vec{\Gamma}_2]&={\rm Tr}_{\rm V}{\rm Tr}_{\rm H}[\vec{G}^{(m)\dagger}\vec{\Gamma}_1\vec{G}^{(m)}\vec{\Gamma}_2]\\
        &= \gamma^2{\rm Tr}_V[(\vec{G}^{(m)}_{1N})^\dagger \vec{G}^{(m)}_{1N}],\\ 
        {\rm Tr}[\vec{G}^{(m)\dagger}\vec{\Gamma}_2\vec{G}^{(m)}\vec{\Gamma}_1] &= \gamma^2{\rm Tr}_{\rm V}[(\vec{G}^{(m)}_{N1})^\dagger \vec{G}^{(m)}_{N1}].
    \end{aligned}
\ee
These components of the Green's function can be easily computed using the recursive algorithm. 

In summary, to enable our simulations we first write all the matrices and equations into the extended space representation; we then perform the recursive calculation using Eqs.~(\ref{eq:floqGreen}) and (\ref{eq:Gmatrix}); from this calculation we obtain the matrix $\boldsymbol{\mathcal{G}}(\omega)$ and extract the Fourier components we need from Eq.~(\ref{eq:Gmatrix}); finally, we assemble those Fourier components according to Eq.~(\ref{eq:FloquetLandauer}) to obtain the differential conductance.  
\end{appendix}
\end{document}